\documentclass[11pt]{article}
\usepackage{graphicx}
\usepackage{amsfonts}

\usepackage[table]{xcolor}

\definecolor{lightgray}{gray}{0.9}

\begin{document}
\title{Random walk hierarchy measure: What is more hierarchical, a chain, a tree or a star?}
\author{D{\'a}niel Cz{\'e}gel$^1$ and Gergely Palla$^{2,3}$\footnote{Corresponding author, e-mail: pallag@hal.elte.hu}\\
\footnotesize{$^1$Dept. of Biological Physics, E{\"o}tv{\"os} University, H-1117 Budapest, Hungary}\\
\footnotesize{$^2$MTA-ELTE Statistical and Biological Physics Research Group,}\\
\footnotesize{Hungarian Academy of Sciences, H-1117 Budapest, Hungary,}\\ 
\footnotesize{$^3$Regional Knowledge Centre, E{\"o}tv{\"os} University, H-8000 Sz{\'e}kesfeh{\'e}rv{\'a}r, Hungary,}}
\maketitle

\begin{abstract}
Signs of hierarchy are prevalent in a wide range of systems in nature and society. One of the key problems is quantifying the importance of hierarchical organisation in the structure of the network representing the interactions or connections between the fundamental units of the studied system. Although a number of notable methods are already available, their vast majority is treating all directed acyclic graphs as already maximally hierarchical. Here we propose a hierarchy measure based on random walks on the network. The novelty of our approach is that directed trees corresponding to multi level pyramidal structures obtain higher hierarchy scores compared to directed chains and directed stars. Furthermore, in the thermodynamic limit the hierarchy measure of regular trees is converging to a well defined limit depending only on the branching number. When applied to real networks, our method is computationally very effective, as the result can be evaluated with arbitrary precision by subsequent multiplications of the transition matrix describing the random walk process. In addition, the tests on real world networks provided very intuitive results, e.g., the trophic levels obtained from our approach on a food web were highly consistent with former results from ecology.
\end{abstract}

\section*{Introduction}

Hierarchical organisation is an ubiquitous feature of a large variety of systems studied in natural- and social sciences. Examples of empirical studies on hierarchy are including the transcriptional regulatory network of Escherichia coli \cite{Zeng_Ecoli}, the dominant-subordinate hierarchy among crayfish \cite{Huber_crayfish}, the leader-follower network of pigeon flocks \cite{Tamas_pigeons,Pigeon_context}, the rhesus macaque kingdoms \cite{McCowan_macaque}, neural networks \cite{Kaiser_neural} and technological networks \cite{Pumain_book}, social interactions \cite{Guimera_hier_soc,our_pref_coms,Sole_hier_soc}, urban planning \cite{Krugman_urban,Batty_urban}, ecological systems \cite{Hirata_eco,Wickens_eco}, and evolution \cite{Eldrege_book,McShea_organism}. Naturally, hierarchy is a very relevant concept also in network theory \cite{Pumain_book,Laci_hier_scale,Newman_hier,Sole_chaos_hier}. The network approach has become an ubiquitous tool for analysing complex systems ranging from the interactions within cells through transportation systems, the Internet and other technological networks to economic networks, collaboration networks and the society \cite{Laci_revmod,Dorog_book}. 

Grasping the signs of hierarchy in networks is a non-trivial task with a number of possible different approaches. On the one hand, we may try the statistical inference of an underlying hierarchy based on the observed network structure, as suggested in Ref.\cite{Newman_hier}. On the other hand, the introduction of a hierarchy measure is also a natural idea \cite{krackhardt1994graph,Sneppen_hier_measures,Luo-Magee_complexity,Enys_hierarchy,Sole_hier_PNAS}. In general, a hierarchy measure, can be viewed as a function on the domain of graphs, $H:\mathbb{G}\mapsto\mathbb{R}$, mapping a graph $\mathcal{G}\in\mathbb{G}$ into a real number, $H(\mathcal{G})\in\mathbb{R}$. The value of the measure is actually $H(\mathcal{G})\in[0,1]$ or $H(\mathcal{G})\in[-1,1]$ in most cases, with high values corresponding to hierarchical structures and low values indicating the absence of hierarchy in the examined network. 

One of the first methods was proposed by D.\ Krackhardt, motivated by organisational hierarchy, defining the hierarchy measure simply as the number of ordered pairs divided by the number of connected pairs \cite{krackhardt1994graph}. In the approach introduced by A.\ Trusina et al., the position of the nodes in the hierarchy is assumed to be given by the degree, and the hierarchy measure is given by the fraction of directed shortest paths going strictly upwards in the hierarchy \cite{Sneppen_hier_measures}. Another way for quantifying the possible asymmetry between nodes is to assume some sort of flow on the links, and examine whether the global map of flows in the system is revealing a kind of overall directionality or not. Probably the simplest approach in this framework is to define the fraction of links not participating in any cycle as the measure of the hierarchy, as suggested by J. Luo and C.\ L.\ Magee \cite{Luo-Magee_complexity}. 

A further important property of a hierarchical system is that reaching the rest of the network should be relatively easy for the nodes high in the hierarchy, and more difficult for the nodes at the bottom of the hierarchy, as pointed out by  E.\ Mones et al.\ in Ref.\cite{Enys_hierarchy}. The hierarchy measure based on this aspect is given by the Global Reaching Centrality, characterising the inhomogeneity of the fraction of reachable nodes in at most $m$-steps in the network \cite{Enys_hierarchy}.  A more elaborate quantification of hierarchy was proposed by B. Corominas-Murta et al.\ in Ref.\cite{Sole_hier_PNAS} with the introduction of Treeness, Feedforwardness and Orderability, projecting the studied network onto a point in a 3 dimensional space, where each dimension is aimed to capture a different aspect of hierarchy. Treeness, $T$, is measuring how ambiguous are the chain of commands in the network, while Feedforwardness, $F$ is related to the size and position of the strongly connected components in the network. Finally, the orderability, $O$ is simply the fraction of nodes not taking part in any directed cycles, i.e., it is analogous to the hierarchy measure introduced by J. Luo and C.\ L.\ Magee. The 3d scatter plots of $T$, $F$ and $O$ provided very interesting results, revealing different clusters of hierarchical networks \cite{Sole_hier_PNAS}. A more detailed description and comparison between the mentioned methods is given in the Supplementary Information S1.

Although the methods listed above allow the examination of the hierarchical organisation from different perspectives, a noteworthy common aspect of these approaches that they all treat acyclic networks as already maximally hierarchical, independent of the further details of the graph structure. (The Global Reaching Centrality given in Ref.\cite{Enys_hierarchy} is an exception, considering the star configuration as the most hierarchical). Here we argue that different acyclic networks are not necessarily equally hierarchical. The general intuition of a hierarchy is usually corresponding to a multi level pyramidal structure, with levels becoming wider and wider as we descend from the root towards the bottom. On the one hand this way the top nodes in the hierarchy can reach most of the network in a very effective way, i.e., via paths of average length scaling as $\ln N$, where $N$ denotes the number of nodes. On the other hand, in this structure all nodes can have a treatable number of direct subordinates. In contrast, if we consider a directed chain, all the levels are of size one, and this is leading to a large average distance scaling as $N$. The other extreme limiting case of acyclic networks is given by the directed star configuration, where all the nodes have a single incoming link from a central hub, and no further out-links. In this case the hierarchy is consisting of only two levels, and the supposed leader in the network has to cope with a number of direct descendants scaling as $N$. Based on that, introducing a hierarchy measure preferring trees to chains and stars would be a substantial step towards achieving a more intuitive approach for evaluating the importance of hierarchy in a network structure. 

In this paper we tackle this problem with the help of random walks on the network. Random walks provide a fundamental model for stochastic processes in a large variety of systems ranging from physics \cite{DeGennes_book}, chemistry \cite{Van_Kampen_book} and computer science \cite{Weiss_book} through biology and ecology \cite{Goel_book,Codling_rw} to economics \cite{Malkiel_book} and psychology \cite{Stone_psy}. In the current problem of quantifying the extent of hierarchy in a network structure, random walkers can be used to evaluate the rank of the nodes in the hierarchy. 

The basic idea is assuming an information flow on the links from nodes high in the hierarchy towards the lower levels, in a similar fashion as in case of a company, where the management is likely to send information and instructions to the employees on a regular basis. Given the network structure, the source of information in the system can be traced back by sending random walkers traversing the links in reverse direction from all nodes. In case the density of the random walkers is reaching a steady state, its value at a given node can be interpreted as the probability that the node was the source of information. Consequently, high random walker density values indicate a high standing in the hierarchy, whereas low density values are corresponding to bottom nodes. The significance of hierarchical organisation in the network structure can be judged based on the inhomogeneity of the distribution: In a homogeneous distribution we cannot pinpoint the source of information, thus, it is corresponding to a non-hierarchical network. In contrast, a very inhomogeneous  distribution is indicating a strongly hierarchical structure.

\section*{Random walk hierarchy measure}
The details of the random walk process are the following. Since the random walkers are traversing the links backwards, the transition probability for a walker from node $j$ to $i$ is proportional to the inverse of the in-degree of $j$, i.e., $P(j\rightarrow i)\propto 1/k_j^{\rm in}$. Another important factor to be taken into account is the limited capacity of the information sources for sending information: In general we can assume that the more out-neighbours a given node has, the less resource it can allocate for managing the communication over a given link. This effect can be taken into account by assuming that $P(j\rightarrow i)$ is also proportional to the out-degree of $i$, i.e., $P(j\rightarrow i)\propto 1/k_{i}^{\rm out}$.Combining the above factors together is resulting in 
\begin{equation}
P(j\rightarrow i)=\frac{1}{k_j^{\rm in}}\frac{1}{k_i^{\rm out}}
\label{eq:trans_k}
\end{equation}
for the transition probability of the random walkers from node $j$ to $i$. (In case $i$ is not an in-neighbour of $j$ the transition probability $P(j\rightarrow i)$  is zero by definition). We note that due to the second factor on the right hand side of (\ref{eq:trans_k}), the probability for staying at the same node can be non-zero in general, given by $P(j\rightarrow j)=1-\sum_{i\neq j} P(j\rightarrow i)$. For weighted networks (\ref{eq:trans_k}) can be naturally generalised to
\begin{equation}
P(j\rightarrow i)=\frac{w_{ij}}{\sum_lw_{lj}}\frac{w_{ij}}{\sum_l w_{il}},
\label{eq:trans_w}
\end{equation}
where $w_{ij}$ denotes the weight of the link from $i$ to $j$.

In case of acyclic networks, all random walkers eventually converge into nodes with no incoming links, (i.e., the ``sources'' in the network). In order to avoid judging the importance of hierarchical organisation in the system solely based on these ``sources'', we inject new random walkers into the network at every time step. The update rules are the following:
\begin{itemize}
\item[1.] We insert $f$ random walkers into the system, increasing the random walker density on every node by $f/N$, thus, the random walker density at node $i$ given by $p_i(t)$ is changing as
\begin{equation}
p_i(t)\mapsto p_i(t)+\frac{f}{N}.
\end{equation}
\item[2.] We let all random walkers in the system proceed on step, governed by the transition probabilities given in (\ref{eq:trans_k}). By introducing a transition matrix $\mathbf{T}$ with matrix elements $T_{ij}=P(j\rightarrow i)$, the density of random walkers on node $i$ after the transition can be expressed as 
\begin{equation}
p_i(t+1)=\sum_{j=1}^N T_{ij}p_j(t).
\end{equation} 
\item[3.] The total sum of random walkers has to be normalised, i.e., we require $\sum_{i=1}^N p_i(t+1)=1$. Since the sum of new random walkers added to the system was $f$, we have to simply divide the density of random walkers by $1+f$ in order to fulfil the normalisation condition,
\begin{equation}
 p_i(t+1)\mapsto \frac{p_i(t+1)}{1+f}.
\end{equation}
\end{itemize}

The above normalisation of the random walker density to unity after each iteration is equivalent to using ``decaying'' random walkers, having a weight decreasing by a factor of $(1+f)^{-1}$ in each step. Let us denote the characteristic distance under which the weight of a random walker is decreased to $e^{-1}$ by $\lambda$, fulfilling 
\begin{equation}
(1+f)^{-\lambda}=e^{-1}.
\end{equation}
Based on that, $f$ can be also expressed as
\begin{equation}
f=e^{1/\lambda}-1.
\end{equation}
Although $\lambda$, (or equivalently, $f$) is a parameter of the method at the current stage, later on in the Results we shall find a natural condition for fixing $\lambda$ at an optimal value in general.

Our main object of interest is the stationary distribution of the random walkers in the network. By writing this distribution in a vector form of $\mathbf{p}^{\rm stat}$, where the $i$-th component of the vector, $p_i^{\rm stat}$, is corresponding to the random walker density on node $i$, we can derive a simple equation based on the update rules. Adding $f/N$ new random walkers at each node is simply incrementing each vector component by $f/N$, while the transition to the next site by the random walkers corresponds to multiplying by the transition matrix $\mathbf{T}$. Finally, the normalisation of the random walker density simply multiplies each vector component by $1/(1+f)$. Based on the above the stationary distribution fulfils 
\begin{equation}
\mathbf{p}^{\rm stat}=\frac{1}{1+f}\left[\mathbf{T}\left(\mathbf{p}^{\rm stat}+\frac{f}{N}\mathbf{1}\right)\right]=e^{-1/\lambda}\left[\mathbf{T}\left(\mathbf{p}^{\rm stat}+\frac{e^{1/\lambda}-1}{N}\mathbf{1}\right)\right],
\label{eq:p_stac_alap}
\end{equation}
where $\mathbf{1}$ is corresponding to a vector of size $N$ with all elements equal to 1. By expressing $\mathbf{p}^{\rm stat}$ we obtain
\begin{equation}
\mathbf{p}^{\rm stat}=\frac{f}{N}\left[(1+f)\mathbf{I}-\mathbf{T}\right]^{-1}\mathbf{T}\mathbf{1}=\frac{e^{1/\lambda}-1}{N}\left[e^{1/\lambda}\mathbf{I}-\mathbf{T}\right]^{-1}\mathbf{T}\mathbf{1},
\label{eq:p_stac}
\end{equation}
where $\mathbf{I}$ is denoting the identity matrix. Since $\mathbf{T}$ is a left stochastic matrix, the absolute value of its largest eigenvalue is 1. Consequently, the absolute value of the eigenvalues of $\frac{1}{(1+f)}\mathbf{T}$ are smaller than 1, and therefore, (\ref{eq:p_stac}) can also be written as
\begin{equation}
\mathbf{p}^{\rm stat}=\frac{f}{N}\sum_{n=1}^{\infty}\left(\frac{1}{1+f}\mathbf{T}\right)^n\mathbf{1}=\frac{e^{1/\lambda}-1}{N}\sum_{n=1}^{\infty}\left(e^{-1/\lambda}\mathbf{T}\right)^n\mathbf{1}.
\label{eq:q}
\end{equation}

This formula is very intuitive, showing explicitly that the stationary distribution of random walkers at a given node is given by the sum of the probabilities of all the paths ending on the node, where the contributions from longer paths are suppressed exponentially as a function of the path length. Based on (\ref{eq:p_stac}-\ref{eq:q}), $\mathbf{p}^{\rm stat}$ can be computed very efficiently. If the size of the network is moderate, we can use (\ref{eq:p_stac}) for obtaining exact results. However, if matrix inversion is becoming computationally expensive, a very good approximation of $\mathbf{p}^{\rm stat}$ can be calculated according to (\ref{eq:q}). I.e., by carrying out the summation up to a certain finite limit $n_{\rm max}$, the obtained result is converging to the exact $\mathbf{p}^{\rm stat}$ exponentially fast.

Our hierarchy measure is based on the inhomogeneity of the stationary distribution of the random walkers. There are several different possibilities for quantifying the inhomogeneity of a probability distribution in general, here we choose the relative standard deviation, (also called as the coefficient of variation). Thus, the random walk hierarchy measure is defined as
\begin{equation}
H=\frac{\sigma(\mathbf{p}^{\rm stat})}{\mu( \mathbf{p}^{\rm stat})},
\end{equation}
where $\mu(\mathbf{p}^{\rm stat})$ and $\sigma(\mathbf{p}^{\rm stat})$ denote the mean and the standard deviation of $p_i^{\rm stat}$ respectively. Since $\sum_{i=1}^N p_i^{\rm stat}=1$, the mean is given simply by $\mu(\mathbf{p}^{\rm stat})=1/N$, and our hierarchy measure can be expressed as
\begin{equation}
H=N\sqrt{\frac{1}{N}\sum_{i=1}^N(p_i^{\rm stat})^2-\frac{1}{N^2}}=
\sqrt{N\sum_{i=1}^N(p_i^{\rm stat})^2-1}.
\label{eq:RWH_final}
\end{equation}

\section*{Results}
\subsection*{Hierarchy of acyclic networks}
For demonstrating the sensitivity of our measure to the topology also in case of acyclic networks, first we evaluate $H$ for chains, regular trees with a constant branching number $b$, and stars. According to calculations detailed in Methods, the corresponding hierarchy values can be expressed as
\begin{eqnarray}
H_{\rm chain}&=&\sqrt{\frac{2}{N}\frac{e^{(3-2N)/\lambda}\left(e^{N/\lambda}-1\right)\left(e^{N/\lambda}-e^{1/\lambda}\right)}{\left(e^{1/\lambda}-1\right)^2\left(e^{1/\lambda}+1\right)}},
\label{eq:RWH_chain_final} \\
H_{\rm tree}&=&\sqrt{N \sum_{l=1}^{l_{\rm max}}b^{l-1}(p_l^{\rm stat})^2-1}, 
\label{eq:RWH_tree_final} \\
H_{\rm star}&=&\frac{\sqrt{(N-1)e^{2/\lambda}}}{(N-1)\left(e^{1/\lambda}-1\right)+1},
\label{eq:RWH_star_final}
\end{eqnarray}
where $N$ is the number of nodes in the networks, $l$ denotes the levels in case of the tree (starting from $l=1$ at the root and ending with $l_{\rm max}$ at the leafs), and $p_l^{\rm stat}$ is corresponding to the stationary distribution of the walkers on level $l$ in (\ref{eq:RWH_tree_final}), which can be obtained from a simple recursion written as
\begin{eqnarray}
p_{l_{\rm max}}^{\rm stat}&=&\frac{e^{1/\lambda}-1}{N}\frac{b-1}{1+be^{1/\lambda}-b},
\label{eq:p_tree_lmax}\\
p_l^{\rm stat}&=&\frac{b}{1+be^{1/\lambda}-b}p_{l+1}^{\rm stat}+\frac{e^{1/\lambda}-1}{N}\frac{2b-1}{1+be^{1/\lambda}-b}
\label{eq:p_tree_l}\\
p_1^{\rm stat}&=&\frac{p_2^{\rm stat}}{e^{1/\lambda}-1}+\frac{2}{N}.
\label{eq:p_tree_first}
\end{eqnarray}

\begin{figure}[h]
\centerline{\includegraphics[width=0.63\textwidth]{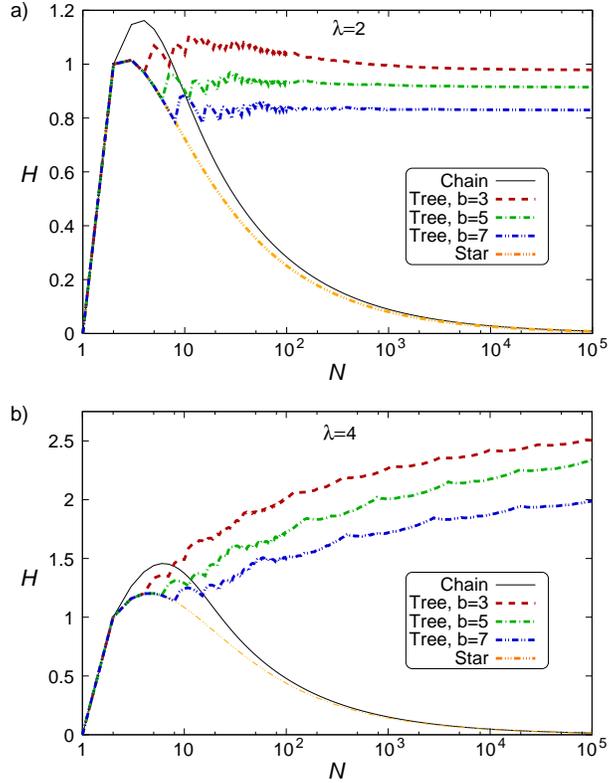}}
\caption{Comparing the random walk hierarchy for chains, regular trees and stars. a) The behaviour of $H$ as a function of $N$ for a chain (black), a regular tree with branching number $b=3$ (red), a regular tree with $b=5$ (green), a regular tree with $b=7$ (blue) and a star (orange) at $\lambda=2$. b) The same plot as in a) when $\lambda$ is set to $\lambda=4$.}
\label{fig:chain_tree_star}
\end{figure}

In Fig.\ref{fig:chain_tree_star}.\ we compare the hierarchy measures given in (\ref{eq:RWH_chain_final}-\ref{eq:RWH_star_final}) at $\lambda=2$ (Fig.\ref{fig:chain_tree_star}a) and at $\lambda=4$ (Fig.\ref{fig:chain_tree_star}b). Our construction algorithm for the trees with a branching number $b$ was to start by adding $b$ links to the root, then move to the second level and subsequently add $b$ links to every node on this level, and so on, move to the next level only when the given level was completed. Whenever the number of nodes in the tree has reached $N$, the algorithm terminates, and naturally, the resulting tree is not completely regular in most of the cases. Nevertheless, the overall structure of the trees obtained in this way is getting closer and closer to regular trees as $N$ is increasing. 

According to Fig.\ref{fig:chain_tree_star}.\ the $H$ for the chain and the star configurations has a peak at very small system sizes, and shows a decreasing tendency for growing $N$. In contrast, for regular trees  $H$ seems more or less converging to a finite value.  Thus, above a certain $N$ it is the structure of the tree, (encoded in the branching number), what determines the hierarchy measure, not the size of the tree. This is indicating that $H$ is behaving similarly to intensive quantities in physics in some aspects. The ``intensive'' property of the hierarchy measure is analysed in more details in the Supplementary Information S2, here we note that if we take a pair of graphs ${\mathcal G}_1$ and ${\mathcal G_2}$  which are not connected to each other, then $H$ for the union of the graphs is equal to the weighted quadratic mean of the $H$ values calculated for the graphs separately,
\begin{equation}
H_{{\mathcal G}_1\cup{\mathcal G}_2}=\sqrt{\frac{N_1\left[H_{{\mathcal G}_1}\right]^2+N_2\left[H_{{\mathcal G}_2}\right]^2}{N_1+N_2}}.
\end{equation}
Thus, in the special case of a pair of isomorphic graphs $H_{{\mathcal G}_1\cup{\mathcal G}_2}=H_{{\mathcal G}_1}=H_{{\mathcal G}_2}$. 

We continue with the examination of the behaviour of $H$ in the thermodynamic limit. According to calculations detailed in Methods, when the system size is diverging, $N\rightarrow\infty$, the hierarchies given in (\ref{eq:RWH_chain_final}-\ref{eq:RWH_star_final}) take the form of
\begin{eqnarray}
H_{\rm chain}&\propto& N^{-1/2},
\label{eq:chain_inf} \\
H_{\rm tree}&=&\left\lbrace \begin{array}{ll}
\sqrt{\frac{(b-1)e^{2/\lambda}}{(1+be^{1/\lambda}-b)^2-b}}, & \lambda< \lambda_{\rm c}(b) \\
\infty & \lambda \geq\lambda_{\rm c}(b)
\end{array}\right. 
\label{eq:RWH_tree_inf} \\
H_{\rm star}&\propto& N^{-1/2}.
\label{eq:RWH_star_inf}
\end{eqnarray}
Thus, the hierarchy measure is vanishing for a chain and a star in the thermodynamic limit. In contrast, $H_{\rm tree}$ is converging to a well defined finite limit value or $b<1<\infty$ when $\lambda$ is smaller than a $b$ dependent critical value, and is diverging otherwise. In the Methods we show that the critical $\lambda$ value is given by
\begin{equation}
\lambda_{\rm c}(b)=\left[\ln\left(\frac{\sqrt{b}-1}{b}+1\right)\right]^{-1}.
\label{eq:lambda_c}
\end{equation}
\begin{figure}[h]
\centerline{\includegraphics[width=0.7\textwidth]{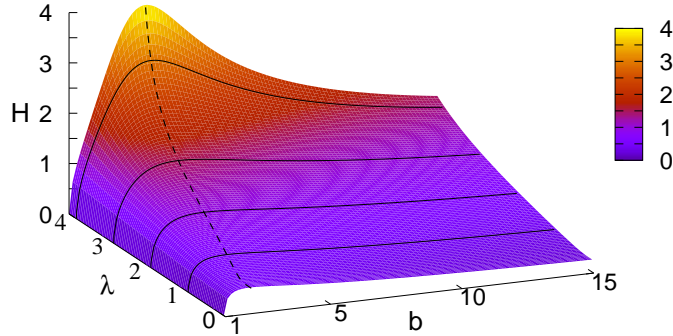}}
\caption{2d plot of the random walk hierarchy measure $H$ for infinitely large regular trees as a function of the branching number $b$ and the parameter $\lambda$. The formula for $H$ is given in (\ref{eq:RWH_tree_inf}). At $b=1$ we recover the infinitely large chain, while the infinitely large star is corresponding to the $b\rightarrow\infty$ limit. The dashed line is showing the maximum place of $H$.}
\label{fig:RWH_inf_trees}
\end{figure}

The behaviour of the limiting $H_{\rm tree}$ given in (\ref{eq:RWH_tree_inf}) is shown in a 2d plot in Fig.\ref{fig:RWH_inf_trees}.\ as a function of $b$ and $\lambda$. At $b=1$ the tree becomes equivalent to an infinitely large chain, and according to (\ref{eq:chain_inf}) the $H$ becomes zero. The 2d surface displayed in Fig.\ref{fig:RWH_inf_trees}.\ is consistent with this result, as it starts from $H=0$ at $b=1$ for all $\lambda$ values. Similarly, the $H$ for an infinitely large star is also zero according to (\ref{eq:RWH_star_inf}). The surface shown in Fig.\ref{fig:RWH_inf_trees}.\ is consistent with this result as well, as we can see a decreasing tendency in $H$ as a function of $b$ in the large $b$ regime. 
In the range of intermediate branching numbers we can observe an $\lambda$ dependent maximum in $H$. 
This behaviour is examined in more details in the Supplementary Information S3.

Based on the behaviour of $H$ in the thermodynamic limit, we can also fix the $\lambda$ parameter at an optimal value in general as follows. Since $\lambda$ is corresponding to the characteristic path length a random walker can traverse before ``decaying'', on the one hand we would like to choose a $\lambda$ as high as possible. I.e., if $\lambda$ is small, the random walkers can explore only within a very limited range from their origin, thus, the information we can retrieve via the random walkers is also very local. However, due to its self similar nature, hierarchical organisation can manifest on all length scales, therefore, we need random walkers travelling longer distances in order to be able to tell apart hierarchical and non-hierarchical networks. 

On the other hand, if $\lambda$ is too large, we may run into diverging hierarchy values according to (\ref{eq:RWH_tree_inf}), which needs to be avoided in case of a well behaving hierarchy measure. Therefore, we fix $\lambda$ at a value as high as possible where a diverging $H$ is avoided for sure even in case of infinitely large regular trees. According to (\ref{eq:lambda_c}), the minimum of $\lambda_{\rm c}(b)$ can be found at $b=4$, where $\lambda_{\rm c}(b=4)\simeq 4.48$. Since the path length traversed by a random walker is increasing by unity under every iteration, it is also natural to set $\lambda$ to an integer value. Based on the above, the optimal setting for $\lambda$ is given by $\lambda=4$. In the rest of the paper we are assuming that $\lambda$ is set to this optimal value, and thereby consider our approach a parameter free method for measuring the amount of hierarchical organisation in the structure of networks.

\begin{figure}[hbt]
\centerline{\includegraphics[width=0.6\textwidth]{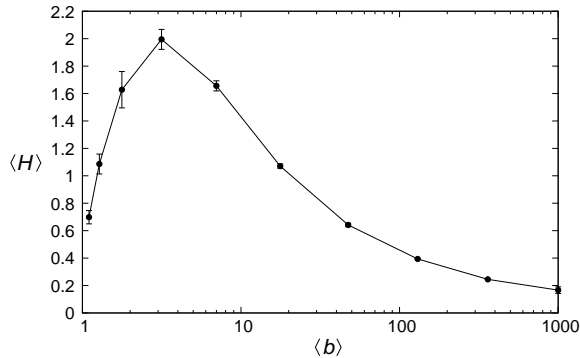}}
\caption{The average random walk hierarchy $\left< H \right>$ as a function of the average branching number $\left< b\right>$ for general trees of $N=1000$ nodes, averaged over $100$ instances. When $\left< b\right>$ is close to one, the tree is basically a chain, whereas at very large branching number, its structure is close to a star.}
\label{fig:Poisson_trees}
\end{figure}

Since real world hierarchies are usually not as highly ordered as a regular tree with a constant branching number, we extended our comparison studies of acyclic graphs also to general directed trees. By applying a simple algorithm detailed in Methods, we generated a large family of trees with branching numbers varying around a given average branching number $\left< b\right>$ according to a shifted Poisson distribution. In Fig.\ref{fig:Poisson_trees}. we show the average of the random walk hierarchy measure, $\left< H\right>$ as a function of $\left< b\right>$, calculated based on 100 realisations of trees consisting of $N=1000$ nodes. According to the curve, the maximum of $\left< H\right>$ is at an intermediate average branching number, where the structure of the network is really tree like. I.e., for low average branching numbers, (where the structure is basically a chain), and also for very large branching numbers comparable to the system size, (where the structure is close to a star), the obtained $\left< H\right>$ values are considerably lower.

\subsection*{Results on real networks}
\subsubsection*{St. Marks food web}
Here we apply our method for analysing the hierarchy of the St.\ Marks food web \cite{Luczkovich_St_Marks}, representing a part of the ecosystem of Goose Creek Bay, St. Marks National Wildlife Refuge, Florida, USA. The nodes of the network are corresponding to living compartments, (group of species) based on probable diet and life history characteristics. Thus, compartments range from single species (e.g., pinfish) through a couple of species (e.g., gulf flounder and needlefish) to large groups of taxa, (e.g., bacterioplankton). The links between the nodes represent the feeding pathways, pointing from consumers to their food sources, where the link weights are corresponding to the fractions of the consumer's diet.   
\begin{figure}[h!]
\centerline{\includegraphics[width=0.95\textwidth]{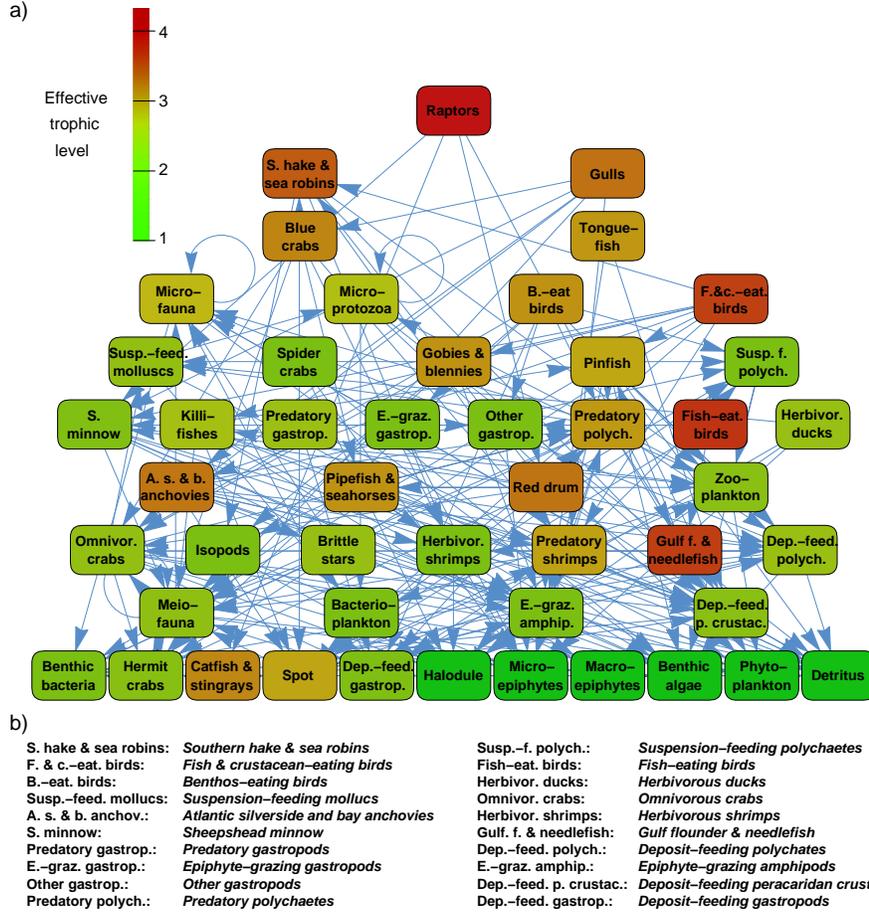}}
\caption{Hierarchy of the St.\ Marks food web. a) The nodes are ordered according to the stationary distribution of the random walkers calculated from (\ref{eq:p_stac}), and the hierarchy levels are corresponding to groups of nodes for which the standard deviation of $p^{\rm stat}_i$ is at most $0.125\cdot\sigma(\mathbf{p}^{\rm stat})$, where $\sigma(\mathbf{p}^{\rm stat})$ denotes the standard deviation of $p_i^{\rm stat}$ over the whole network. The colour coding of the nodes reflects their effective trophic level published in \cite{Luczkovich_St_Marks}.  b) Listing of the abbreviations used in a).}
\label{fig:St_Marks}
\end{figure}

The static distribution of the random walkers on the network defined above can be calculated using (\ref{eq:p_stac}). However, $p^{\rm stat}_i$ is defining only a ranking between the nodes and does not provide the hierarchy levels in the first place. Therefore, we sampled and aggregated nodes into levels so that in each level, the standard deviation of $p^{\rm stat}_i$ is lower than a pre-defined fraction of the standard deviation in the whole network. (This type of procedure for obtaining the hierarchy levels was established in \cite{Enys_hierarchy}).

 In Fig.\ref{fig:St_Marks}.\ we show the resulting hierarchy between the compartments when the standard deviation of $p_i^{\rm stat}$ within the levels is at most $0.125\cdot \sigma(\mathbf{p}^{\rm stat})$. The hierarchy levels are consistent with the common sense about food webs as e.g., benthic algae is on the lowest level, herbivorous ducks are somewhere in the middle, and raptors (such as e.g., the bald eagle) are on the top of the hierarchy. The colour coding of the nodes is showing the effective trophic level of the compartments given in \cite{Luczkovich_St_Marks}, ranging between 1.0 and 4.32. Apparently, the position of the nodes in the hierarchy and their colour are coherent in most of the cases, e.g., the root has the highest effective trophic level, and the nodes with the lowest trophic level are at the bottom of the hierarchy. However, a small number of discrepancies can be also observed, (e.g., as in case of Gulf flounder \& needlefish), signing that the effective trophic levels and the random walk based hierarchy are catching slightly different aspects of the studied food web. 

Finally, the Spearman's rank correlation coefficient between the ranking of the compartments according to $p_i^{\rm stat}$ and the ranking according to the effective trophic levels is $0.593$. In contrast, the Spearman's rank correlation coefficient between the effective trophic levels and the hierarchy obtained after applying a degree preserving link randomisation to the network is only $0.006\pm 0.138$. Based on the above, our hierarchy is highly consistent with former results from ecology.

\subsection*{Comparing different networks}
We also calculated the $H$ given in (\ref{eq:RWH_final}) for numerous different systems ranging from metabolic and regulatory networks through citation, trust and language networks to the Internet and the WWW. (A detailed description of the networks is given in the Supplementary Information S4). In Fig.\ref{fig:scatter}.\ we show  the obtained hierarchy values as a function of the network size, $N$. 
\begin{figure}[hbt]
\centerline{\includegraphics[width=0.7\textwidth]{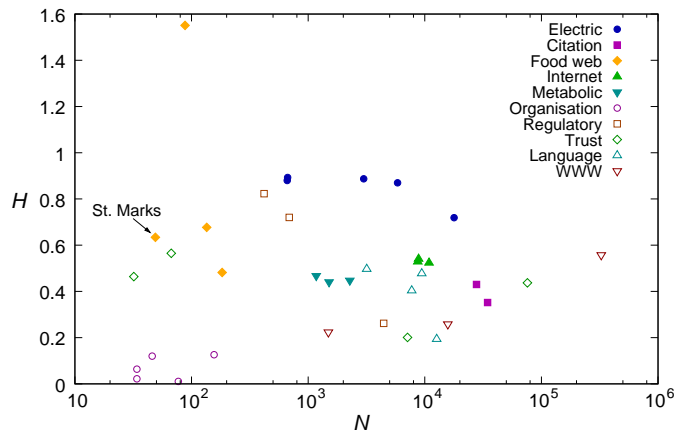}}
\caption{Random walk hierarchy of different real networks. Each symbol is corresponding to a different network, where the shape and colour of the symbols is encoding the type of the system. The horizontal coordinate of the symbols is corresponding to the size of the corresponding network, while the vertical coordinate is giving $H$.}
\label{fig:scatter}
\end{figure}
According to the figure food webs, electric circuits and regulatory networks provide the largest $H$ values, and in contrast, the informal networks of acquaintances in different organisations seem the least hierarchical. In the mean time, the WWW, the Internet, the citation-, metabolic-, trust- and language networks appear to be moderately hierarchical.

However, under certain circumstances we may obtain a moderate hierarchy measure even in a random graph. E.g., the structure of the giant component in the Erd\H os--R\'enyi graph \cite{Erdos} is more or less tree-like if we are close to the percolation threshold, and tree-like structures are usually considered highly hierarchical. Accordingly, in order to make a fair judgement on the importance of hierarchy in the topology of a real network, we need to compare the measured $H$ to the result we expect in a suitably chosen random network ensemble, modelling the structure of the given network under the assumption of random connections. In order to take into account of the degree distribution of the studied networks, we use the configuration model for evaluating the expected value of $H$ in the random network ensemble. A sample from this ensemble can be obtained by simply link randomising the given real network, keeping the degree of the nodes fixed under the random rewiring of the connections. 

The difference between $H$ obtained for the real networks and the expected value of $H$ in their random counterparts can be measured in terms of the $z$-score, defined as
\begin{equation}
z=\frac{H-\left< H\right>}{\sigma(H)},
\end{equation}
where $\left< H\right>$ and $\sigma(H)$ denote the expected value- and the standard deviation of $H$ in the random ensemble, respectively. Thus, we basically scale the difference between the real $H$ and the average of $H$ over the random ensemble by the standard deviation of $H$ in the random ensemble. 

\begin{figure}[hbt]
\centerline{\includegraphics[width=0.67\textwidth]{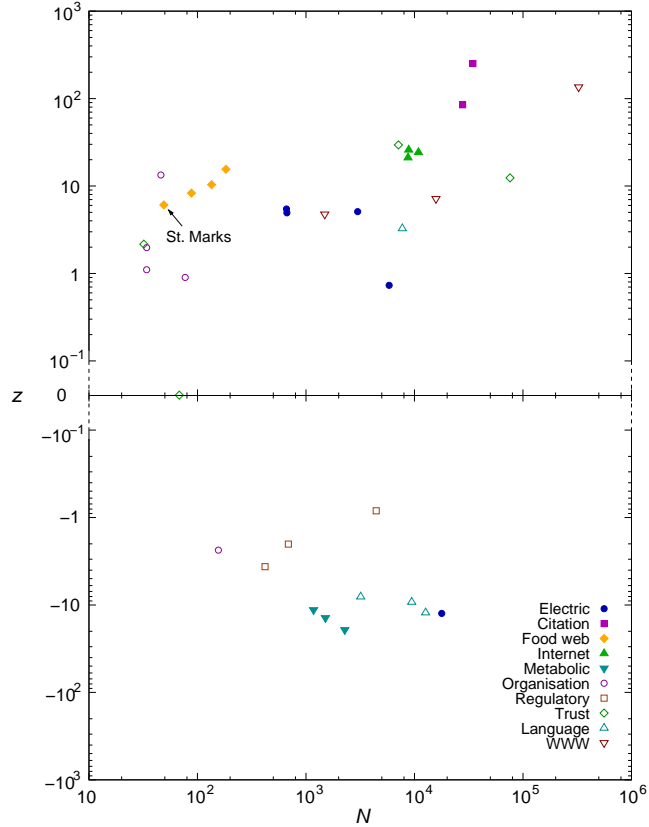}}
\caption{The $z$-score of $H$ for the networks shown in Fig.\ref{fig:scatter}. The $z$-score, given by $z=(H-\left< H\right>)/\sigma(H)$ is plotted as a function of the system size $N$. The random null-model for evaluating $\left< H\right>$ and $\sigma(H)$ is corresponding to the configuration model.}
\label{fig:z_score}
\end{figure}

In Fig.\ref{fig:z_score}. we show the $z$-scores corresponding to the $H$ values displayed in Fig.\ref{fig:scatter}. According to the results, the citation networks and the network between the web pages of the nd.edu domain achieve outstandingly high $z$-scores. Furthermore, all of the food webs, the Internet networks and also the rest of the WWW networks obtain considerably large positive $z$-scores. This means that the structure of these networks is far more hierarchical compared to a random network with the same degree distribution. In contrast, all of the regulatory- and metabolic networks have negative $z$-scores, (with rather large absolute values in the latter case). Thus, these networks are less hierarchical compared to what we would expect on a random base.
Finally, in case of the electric-, organisational-, and language networks we see a mixed picture, where both positive and negative $z$-scores occur. Most of the organisational networks have positive $z$-scores, reaching to a quite high value in case Consulting network, while in parallel we obtain a negative $z$-score for the Enron network. The word adjacency network for the French, Spanish and Japanese languages have negative $z$-scores, opposed to a clearly positive $z$-score in case of the English language. A more detailed analysis of these results is provided in the Supplementary Information S4.

\section*{Discussion}
Measuring the significance of hierarchical organisation in the structure of a complex network is a non-trivial problem with a number of different options available. Here we have proposed a novel method based on random walks on the network. The basic idea behind our approach is that if nodes were sending instructions or information over the links to their subordinates, then the sources of the information could be traced using random walkers traversing the links backwards. The update rules of the dynamics are chosen in a way to make the density of the walkers on the nodes converge to a stationary distribution exponentially fast with the number of iterations. The position of the nodes in the hierarchy is determined by this distribution, with high random walker densities corresponding to top nodes, and low values of the distribution signalling bottom nodes. The overall measure of the hierarchy is given by the inhomogeneity of the stationary distribution. 

The calculation of the hierarchy measure can be carried out based on repeated multiplications of an $N$ by $N$ transition matrix, making the method computationally very efficient and opens up the possibility for GPU based parallelisation. The other main advantage of our approach is that it can differentiate between directed acyclic graphs of distinct nature, opposed to most other methods treating all directed acyclic networks as already maximally hierarchical. The random walk hierarchy measure provides higher scores for trees showing a multilevel pyramidal structure compared to chains and stars. This is consistent with a general intuitive picture about hierarchies: the multilevel pyramidal structure enables the leaders in the tip of the hierarchy to reach the rest of the system via relatively short paths, and also avoids the ``overloading'' of any nodes with a too large number of direct subordinates. In contrast, the distance between the top and the bottom becomes very large in a chain, while the number of direct subordinates for the central node is diverging with the system size in case of a star. A further interesting property of our measure is that it is behaving similarly to intensive quantities in physics. I.e., for regular trees with a constant branching number the hierarchy measure is converging to a well defined value in the thermodynamic limit. Thus, above a certain scale it is the structure, (encoded in the branching number), what determines the hierarchy, not the size of the network. 

Moreover, our tests on real world networks provided rather encouraging results. On the one hand, the detailed analysis of the St.\ Marks food web resulted in hierarchy levels that are highly consistent with former results from ecology on the effective trophic levels in the system. On the other hand, the large scale analysis of numerous further real networks revealed that the value of the hierarchy measure on its own does not always provide a fair characterisation of the importance of hierarchy in the structure of the studied system. According to our results, in some cases a relatively low $H$ value can be accompanied by an outstandingly high $z$-score, when we compare the actual $H$ to the expected value of $H$ in a randomly rewired network with the same degree distribution. This leads to the conclusion that the basic network characteristics such as the link density, degree distribution, etc. can inflict some constrains on the possible range of $H$ and also on $\left< H\right>$ in the corresponding random network ensemble. However, the further analysis of these effects is out of the scope of the present paper and is providing interesting directions for further research.

\section*{Methods}

\subsection*{Hierarchy of chains, regular trees and stars}
First we note that $H=0$ when the network is consisting of only a single directed cycle. Since all the nodes are equivalent in this case, $p_i^{\rm stat}=1/N$ for all $i$, thus, $\sigma(\mathbf{p}^{\rm stat})=0$. Now let us examine how does $H$ change if we move from a cycle to a chain by cutting a single link. Since all the nodes have a unit in-degree except the first node, and all the nodes have a unit out-degree except the last node, the transition probability from level $l$ to level $l-1$ is unity. Therefore, the stationary distribution on the last node is zero, $p_N^{\rm stat}=0$, (since the total amount of injected random walkers exit immediately, and there is no inflow of walkers from outside). Based on (\ref{eq:p_stac_alap}), the stationary distribution of the walkers on the intermediate levels fulfils 
\begin{equation}
p_l^{\rm stat}=\frac{1}{1+f}\left( p_{l+1}^{\rm stat}+\frac{f}{N}\right)=e^{-1/\lambda}\left(p_{l+1}^{\rm stat}+\frac{e^{1/\lambda}-1}{N}\right),
\label{eq:chain}
\end{equation}
while in case of the first node the injected random walkers cannot exit the node, resulting in 
\begin{equation}
p_1^{\rm stat}=\frac{1}{1+f}\left[p_2^{\rm stat}+\frac{f}{N}+p_1^{\rm stat}+\frac{f}{N}\right]=e^{-1/\lambda}\left[p_2^{\rm stat}+p_1^{\rm stat}+2\frac{e^{1/\lambda}-1}{N}\right]
\label{eq:chain_1}
\end{equation}
By solving (\ref{eq:chain}-\ref{eq:chain_1}) we gain
\begin{eqnarray}
p_l^{\rm stat}&=&\frac{1}{N}\left[1-\frac{1}{(1+f)^{N-l}}\right]=\frac{1-e^{-\frac{N-l}{\lambda}}}{N}
 \;\;\; (2\leq l\leq N), \\
p_1^{\rm stat}&=&\frac{p_2^{\rm stat}}{f}+\frac{2}{N}=\frac{p_2^{\rm stat}}{e^{1/\lambda}-1}+\frac{2}{N}.
\end{eqnarray}
By substituting into (\ref{eq:RWH_final}) the result simplifies to (\ref{eq:RWH_chain_final}).

The random walk hierarchy measure for general trees with varying branching number cannot be given in a general formula, nevertheless it can be calculated exactly for any particular finite tree based on (\ref{eq:p_stac}) and (\ref{eq:RWH_final}). However, in case of a regular tree with branching number $b$, a simple recursion can be given for the stationary distribution of the random walkers, as the transition probability from any node to its ``leader'' in the level above is simply $1/b$. The random walkers cannot exit from the root, which we label as level $l=1$, and there is an inflow of walkers from the second level, resulting in
\begin{eqnarray}
p_1^{\rm stat}&=&\frac{1}{1+f}\left[b\left(p_2^{\rm stat}+\frac{f}{N}\right)\frac{1}{b}+p_1^{\rm stat}+\frac{f}{N}\right]= \nonumber \\
 &=&e^{-1/\lambda}\left[p_2^{\rm stat}+p_1^{\rm stat}+2\frac{e^{1/\lambda}-1}{N}\right],
\label{eq:p_tree_1}
\end{eqnarray}
which is exactly the same as in case of the first node in the chain, given in (\ref{eq:chain_1}). For the intermediate levels, we have an inflow of walkers from the level below, and also a term corresponding to the probability of the walkers staying at the given level instead of moving to the level above, yielding altogether
\begin{eqnarray}
p_l^{\rm stat}&=&\frac{1}{1+f}\left[ b\left(p_{l+1}^{\rm stat}+\frac{f}{N}\right)\frac{1}{b}+\left(p_l^{\rm stat}+\frac{f}{N}\right)\left(1-\frac{1}{b}\right)\right]= \nonumber \\
& &e^{-1/\lambda}\left[p_{l+1}^{\rm stat}+\frac{e^{1/\lambda}-1}{N}+\left(p_l^{\rm stat}+\frac{e^{1/\lambda}-1}{N}\right)\left(1-\frac{1}{b}\right)\right]
\label{eq:p_tree_l_meth}
\end{eqnarray}
Finally, on the last level $l=l_{\rm max}$ we have no inflow from other nodes, giving
\begin{eqnarray}
p_{l_{\rm max}}^{\rm stat}&=&\frac{1}{1+f}\left(p_{l_{\rm max}}^{\rm stat}+\frac{f}{N}\right)\left(1-\frac{1}{b}\right)=\nonumber \\
& &e^{-1/\lambda}\left(p_{l_{\rm max}}^{\rm stat}+\frac{e^{1/\lambda}-1}{N}\right)\left(1-\frac{1}{b}\right),
\end{eqnarray} 
which provides an immediate solution for $p_{l_{\rm max}}$ in the form given in (\ref{eq:p_tree_lmax}). Based on (\ref{eq:p_tree_lmax}) we can calculate the stationary distribution on the rest of the levels as well, i.e., by rearranging (\ref{eq:p_tree_1}) and (\ref{eq:p_tree_l_meth}) we gain  (\ref{eq:p_tree_l}) and (\ref{eq:p_tree_first}).

The hierarchy measure for the star can be evaluated in a similar fashion to that of the chain. In this case $N-1$ peripheral nodes are connected to a central node, from which the random walkers cannot exit.  Thus, the stationary distribution of the random walkers fulfil
\begin{eqnarray}
p_{\rm c}^{\rm stat}&=&\frac{1}{1+f}\left[p_{\rm c}^{\rm stat}+\frac{f}{N}
+(N-1)\left(p_{\rm p}^{\rm stat}+\frac{f}{N}\right)\frac{1}{N-1}\right]=\nonumber \\
& &e^{-1/\lambda}\left[p_{\rm c}^{\rm stat}+p_{\rm p}^{\rm stat}+2\frac{e^{1/\lambda}-1}{N}\right],
\label{eq:star_cent}\\
p_{\rm p}^{\rm stat}&=&\frac{1}{1+f}\left(p_{\rm p}^{\rm stat}+\frac{f}{N}\right)\left(1-\frac{1}{N-1}\right)=\nonumber \\
& &e^{-1/\lambda}\left(p_{\rm p}^{\rm stat}+\frac{e^{1/\lambda}-1}{N}\right)\left(1-\frac{1}{N-1}\right)
\label{eq:star_peri}
\end{eqnarray}
where $p_{\rm c}^{\rm stat}$ denotes the density on the central node, and $p_{\rm p}^{\rm stat}$ is equal to the density on the peripheral nodes. From (\ref{eq:star_peri}) we can express $p_{\rm p}^{\rm stat}$ directly as
\begin{equation}
p_{\rm p}^{\rm stat}=\frac{f}{N}\frac{N-2}{f(N-1)+1}=\frac{e^{1/\lambda}-1}{N}\frac{N-2}{(e^{1/\lambda}-1)(N-1)+1},
\label{eq:star_p_res}
\end{equation}
and by substituting (\ref{eq:star_p_res}) into (\ref{eq:star_cent}) we arrive to
\begin{equation}
p_{\rm c}^{\rm stat}=\frac{1}{N}\left(2+\frac{N-2}{f(N-1)+1}\right)=\frac{1}{N}\left(2+\frac{N-2}{(e^{1/\lambda}-1)(N-1)+1}\right)
\label{eq:star_c_res}
\end{equation}
for the central node. According to (\ref{eq:RWH_final}), the random walk hierarchy measure can be given in this case as
\begin{equation}
H_{\rm star}=\sqrt{N\left[(p_c^{\rm stat})^2+(N-1)(p_p^{\rm stat})^2\right]-1}.
\label{eq:RWH_star_alap}
\end{equation}
By substituting (\ref{eq:star_p_res}) and (\ref{eq:star_c_res}) into (\ref{eq:RWH_star_alap}), the resulting formula can be simplified to (\ref{eq:RWH_star_final}).

\subsection*{Hierarchy in the thermodynamic limit}
Taking the $N\rightarrow\infty$ limit of $H_{\rm chain}$ given in (\ref{eq:RWH_chain_final}) and of $H_{\rm star}$ written in (\ref{eq:RWH_star_final}) is trivial, the results are given in (\ref{eq:chain_inf}) and in (\ref{eq:RWH_star_inf}) respectively. In contrast, the evaluating the $N\rightarrow\infty$ limit of $H_{\rm tree}$ given in (\ref{eq:RWH_tree_final}) is more complicated and can be carried out as follows. 

First we separate the first term from the rest in the sum over the levels in (\ref{eq:RWH_tree_final}) as
\begin{equation}
H_{\rm tree}=\sqrt{N \left(\sum_{l=2}^{l_{\rm max}}b^{l-1}(p_l^{\rm stat})^2+(p_1^{\rm stat})^2\right)-1}.
\label{eq:sum_sep}
\end{equation} 
In order to evaluate the remaining sum, we express $p_l^{\rm stat}$ given in (\ref{eq:p_tree_l}) as
\begin{equation}
p_l^{\rm stat}=\frac{1}{N}\left[A+B\cdot C^{l_{\rm max}-l+1}\right],
\end{equation}
where
\begin{equation}
A=\frac{(2b-1)f}{1+b(f-1)},~~~
B=-\frac{bf(1+f)}{1+b(f-1)},~~~
C=\frac{b}{1+bf}.
\label{eq:letter_def}
\end{equation}
Based on the above,
\begin{eqnarray}
& &N\left(\sum_{l=2}^{l_{\rm max}} b^{l-1}(p_l^{\rm stat})^2\right)=
\frac{1}{N}\sum_{l=2}^{l_{\rm max}} b^{l-1}\left[A+BC^{l_{\rm max}-l+1}\right]^2= 
\nonumber \\ 
& &
\frac{1}{N}\sum_{l=2}^{l_{\rm max }}\left[A^2 b^{l-1} + 2 AB b^{l-1}C^{l_{\rm max}-l+1} + B^2 b^{l-1} C^{2(l_{\rm max}-l+1)}\right]= \nonumber \\ 
& &\frac{1}{N}\left\{A^2\left(\frac{b^{l_{\rm max}}-1}{b-1}-1\right) + 2 AB b^{l_{\rm max}}\left[\frac{\left(\frac{C}{b}\right)^{l_{\rm max}-l+1}}{\frac{C}{b}-1}-1\right]+  \right. \nonumber \\ & &\left. +B^2 b^{l_{\rm max}}\left[\frac{\left(\frac{C^2}{b}\right)^{l_{\rm max}-l+1}}{\frac{C^2}{b}-1}-1\right]\right\}. \label{eq:sum_temp}
\end{eqnarray}
By using that $N=\frac{b^{l_{\rm max}}-1}{b-1}\approx\frac{b^{l_{\rm max}}}{b-1}$ if $N>>1$, (\ref{eq:sum_temp}) can be also written as
\begin{eqnarray}
& &N\left(\sum_{l=2}^{l_{\rm max}} b^{l-1}(p_l^{\rm stat})^2\right)=\frac{1}{N}\bigg\{A^2(N-1)-(2AB+B^2)\left[N(b-1)+1\right]+ \nonumber \\ & & -\frac{2AB}{C/b-1}\left[N(b-1)+1\right]-\frac{B^2}{C^2/b-1}\left[N(b-1)+1\right]\bigg\}+  \nonumber \\ & & + \frac{b-1}{b^{l_{\rm max}}}\left\{\frac{2AB}{C/b-1}C^{l_{\rm max}}+\frac{B^2}{C^2/b-1}C^{2l_{\rm max}}\right\}=\nonumber \\ & &
A^2-(2AB+B^2)(b-1)-\frac{2AB}{C/b-1}(b-1)-\frac{B^2}{C^2/b-1}(b-1)+\nonumber \\ & &+\mathcal{O}\left(\frac{1}{N}\right)+(b-1)\left[\frac{2AB}{C/b-1}\left(\frac{C}{b}\right)^{l_{\rm max}}+\frac{B^2}{C^2/b-1}\left(\frac{C^2}{b}\right)^{l_{\rm max}}\right]
\label{eq:sum_temp2}
\end{eqnarray}

According to (\ref{eq:letter_def}) 
\begin{equation}
\frac{C}{b}=\frac{1}{1+bf}<1.
\end{equation}
However, the similar inequality of 
\begin{equation}
\frac{C^2}{b}=\frac{b}{(1+bf)^2}<1
\end{equation}
holds if and only 
\begin{equation}
\frac{\sqrt{b}-1}{b}<f,
\label{eq:b_cond}
\end{equation}
or in terms of $\lambda$ if and and only
\begin{equation}
\lambda<\left[ \ln\left(\frac{\sqrt{b}-1}{b}+1\right)\right]^{-1}.
\label{eq:lam_cond}
\end{equation}
Thus, when (\ref{eq:b_cond}), or equivalently (\ref{eq:lam_cond}) are fulfilled, the last two terms in (\ref{eq:sum_temp2}) vanish if $l_{\rm max}\to\infty$. By using (\ref{eq:letter_def}) and neglecting the $\mathcal{O}\left(\frac{1}{N}\right)$ terms we obtain
\begin{equation}
\lim_{l_{\rm max}\to\infty} N\left(\sum_{l=2}^{l_{\rm max}} b^{l-1}(p_l^{\rm stat})^2\right)=
\frac{f(b^2f+b(4+f)-f-2)}{b^2f^2+b(2f-1)+1}.
 \label{eq:sum_fin}
\end{equation}

Now let us examine the $(p_1^{\rm stat})^2$ term in (\ref{eq:sum_sep}). According to (\ref{eq:p_tree_1}) we can write
\begin{equation}
p_{1}^{\rm stat}=\frac{p_{2}^{\rm stat}}{f}+\mathcal{O}\left(\frac{1}{N}\right),
\end{equation}
where 
\begin{equation}
p_{2}^{\rm stat}=\frac{b-1}{b^{l_{\rm max}-1}}\left(A+BC^{l_{\rm max}-1}\right)=\frac{(b-1)B}{C}\left(\frac{C}{b}\right)^{l_{\rm max}}+\mathcal{O}\left(\frac{1}{N}\right).
\end{equation}
Since $l_{\rm max}=\mathcal{O}(\log N)$ and $\frac{C}{b}<1$,
\begin{equation}
\left(\frac{C}{b}\right)^{l_{\rm max}}\propto e^{-\log N}=\frac{1}{N},
\end{equation}
and we obtain that
\begin{equation}
p_1^{\rm stat}=\mathcal{O}\left(\frac{1}{N}\right).
\end{equation}
As a consequence, $N (p_1^{\rm stat})^2=\mathcal{O}\left(\frac{1}{N}\right)$, which is also vanishing when $N\to\infty$. Hence, by substituting (\ref{eq:sum_fin}) into (\ref{eq:sum_sep}) and neglecting the $\mathcal{O}\left(\frac{1}{N}\right)$ terms we arrive to 
\begin{eqnarray}
\lim_{l_{\rm max}\rightarrow \infty} H_{\rm tree}&=&\lim_{l_{\rm max}\to\infty} \sqrt{N\left(\sum_{l=2}^{l_{\rm max}}b^{l-1}(p_l^{\rm stat})^2+(p_{1}^{\rm stat})^2\right)-1}=\nonumber \\
& &\sqrt{\frac{(b-1)(f+1)^2}{b^2f^2+b(2f-1)+1}}=\sqrt{\frac{(b-1)e^{2/\lambda}}{(1+be^{1/\lambda}-b)^2-b}},
\end{eqnarray}
equivalent to the formula given in (\ref{eq:RWH_tree_inf}).

\subsection*{Generating trees with a varying branching number}
We used the following algorithm for generating a tree with varying branching number between $N$ nodes:
\begin{enumerate}
\item[(i)] Initially the nodes are ordered, however, they are also completely isolated from each other
\item[(ii)] We iterate over the nodes according to their order. For current node $i$ we draw a number $\kappa(i)$ from a Poisson distribution with a fixed parameter $\alpha$, and assign the branching number $b(i)=\kappa(i)+1$ to the node. (This way it is guaranteed that the branching number of $i$ is $b(i)\geq 1$).
\item[(iii)] We scan the nodes coming after $i$ and stop at the first node $j$ with no incoming link. We attach directed links pointing from $i$ to the nodes starting from $j$ and ending at $j+b(i)-1$.
\item[(iv)] We repeat steps (ii)-(iii) until all nodes become connected to the tree.
\end{enumerate}
The advantage of this algorithm is that it enables the study of the swift transition from a chain through a family of trees with increasing average branching number to a star. I.e., if we set the parameter of the Poisson distribution to $\alpha=0$, we obtain $b(i)=1$ for all nodes, thus, the resulting graph is actually a chain. However, if $\alpha$ is large enough compared to $N$, the branching number drawn for the first node is already larger than $N$, thus, we obtain a star. For intermediate parameter values the average branching number of the tree is of course $\left< b\right>=1+\alpha$. However, the branching numbers of the individual nodes in the tree will deviate from this average in a similar manner to real systems.

\section*{Acknowledgements}
The authors are grateful to Tam{\'a}s Vicsek and Enys Mones for the fruitful discussions and help in the data collection. The research was partially supported by the European Union and the European Social Fund through project FuturICT.hu (grant no.:TAMOP-4.2.2.C-11/1/KONV-2012-0013) and by the Hungarian National Science Fund (OTKA K105447). The funders had no role in study design, data collection and analysis, decision to publish, or preparation of the manuscript.

\section*{Author contributions}
GP developed the concept of the study. DC and GP worked out the definition of the hierarchy measure, DC derived the behaviour in the thermodynamic limit. DC collected the network data and carried out its analysis. DC and GP prepared the figures and contributed to the result interpretation. GP drafted the manuscript.

\section*{Additional information}
Supplementary information accompanies this paper.\\
The authors declare no competing financial interest.


\newpage

\begin{center}
\Huge{\bf Supplementary Information}
\end{center}
\vspace{1cm}

\renewcommand{\thefigure}{S\arabic{figure}}
\renewcommand{\thetable}{S\arabic{table}}
\renewcommand{\theequation}{S\arabic{equation}}
\renewcommand{\thesection}{S\arabic{section}}
\setcounter{equation}{0}
\setcounter{figure}{0}

\section{Measuring the level of hierarchical organisation in a network}
Characterising the importance of hierarchy in a given network structure is a non-trivial problem with a large number of alternative approaches. Without loss of generality we can formulate a few intuitive requirements a hierarchy measure should meet: First, we assume no a priori ordering between the nodes, the measure is evaluated purely based on the topology of the network. Moreover, the hierarchy measure should should not be too sensitive to the local structure of the network, i.e., the degree sequence alone should not provide enough information for complete evaluation of $H$. In other words, we should be able to modify $H$ when rewiring the links in a network without changing the degree sequence. E.g., let us suppose that the network is corresponding to a single directed cycle of $N$ nodes, (a giant directed ``ring''). Since all the nodes are equivalent, the network lacks any hierarchy what so ever, thus, $H$ should be equal to 0. However, if we take just a single link away, (which is only a minor change if $N$ is large), the network becomes a directed chain, which is indeed hierarchical, thus, a significant jump should be observed in $H$. 

One of the first hierarchy measure was proposed by D.\ Krackhardt, motivated by organisational hierarchy \cite{krackhardt1994graph}. The main assumption here is that in a hierarchical company we can reach the lower levels of the hierarchy from the levels above via chains of commands, and in contrast, we cannot reach the higher levels in a similar fashion from the levels below. Thus, on the one hand, we count the total number of ordered pairs $(i,j)$ in the network for which there is directed path either from $i$ to $j$ or from $j$ to $i$, but not both. On the other hand, we also evaluate the total number of connected pairs, for which at least one directed path exists between the two nodes, (but we also allow paths in both directions). By denoting the number of ordered pairs by $A$, and the number of connected pairs by $C$, Krackhardt's hierarchy measure \cite{krackhardt1994graph} is simply given by
\begin{equation}
H_{\rm K}=\frac{A}{C}.
\end{equation}

Based on the definition above, $H_{K}=1$ for all acyclic networks, where all the node pairs are ordered. Nevertheless, $H_{K}$ is a fine example for a hierarchy measure depending mostly on the global structure of the network: For a directed cycle of $N$ nodes $H_{K}=0$, (since we can reach any node from any other node, and hence, every pair is unordered). By deleting a single link we turn the network into a directed chain, (where all the pairs are ordered), and suddenly the hierarchy measure jumps to $H_{K}=1$.

An approach motivated by similar intuitions compared to Krackhard's hierarchy was introduced by E.\ Mones et al., assuming that reaching the rest of the network should be relatively easy for the nodes high in the hierarchy, and more difficult for the nodes at the bottom of the hierarchy \cite{Enys_hierarchy}. Here the position of the node $i$ in the hierarchy is determined by its reaching centrality, $C_{\rm R}(i)$, corresponding to the fraction of nodes that can be reached from $i$, (following directed paths). Based on the $C_{\rm R}$ associated to the individual nodes, the Global Reaching Centrality of the whole network is defined as \cite{Enys_hierarchy}
\begin{equation}
H_{\rm GRC}=\frac{\sum_{i=1}^N \left[C_{\rm R}^{\rm max}-C_{\rm R}(i)\right]}{N-1},
\label{eq:GRC}
\end{equation}
where the summation is running over the nodes, the size of the network is given by $N$ and the maximal reaching centrality is denoted by $C_{\rm R}^{\rm max}$. 

The maximal possible value of this hierarchy measure is $H_{\rm GRC}=1$, which is obtained when the network is corresponding to a star with $N-1$ arms, with only the central node having a non zero reaching centrality. Interestingly, for a chain of $N$ nodes, we obtain $H_{\rm GRC}=1/2\cdot N/(N-1)$, which is still larger than $1/2$. Similarly to the previous measure, $H_{\rm GRC}=0$ for a directed cycle. 

Another way for quantifying the possible asymmetry between nodes is to assume some sort of flow on the links, and examine whether the global map of flows in the system is revealing a kind of overall directionality or not. Probably the simplest approach in this framework is to define the fraction of links not participating in any cycle as the measure of the hierarchy. I.e., the link flow hierarchy proposed by J. Luo and C.\ L.\ Magee can be formulated as \cite{Luo-Magee_complexity}
\begin{equation}
H_{\rm LF}=\frac{M_{\rm ac}}{M},
\label{eq:linkflow}
\end{equation}
where $M_{\rm ac}$ denotes the number of acyclic links, not part of any directed cycle, and $M$ is corresponding to the total number of links in the network.

Similarly to Krackhardt's hierarchy, the link flow hierarchy is $H_{\rm LF}=1$ for all acyclic networks. Furthermore, when the network is corresponding to a single directed cycle, $H_{\rm LF}=0$. Thus, when deleting a link from this cycle, we observe a jump from zero hierarchy to maximal hierarchy.

A more elaborate quantification of hierarchy was proposed by B. Corominas-Murta et al. \cite{Sole_hier_PNAS} with the introduction of treeness, feedforwardness and orderability, projecting the studied network onto a point in a 3 dimensional space, where each dimension is aimed to capture a different aspect of hierarchy. Treeness, $T$, is measuring how ambiguous are the chain of commands in the network. I.e., in a regular tree where links are pointing from higher levers to lower levels we obtain $T=1$, whereas if revert the link directions, the obtained structure is considered anti-hierarchical, with $T=-1$. The calculation of $T$ is based on comparing forward and backward entropies \cite{Sole_hier_PNAS}. 

In the mean time Feedforwardness, $F$ is related to the size and position of the strongly connected components in the network. Since we can reach any node from any other node in a strongly connected component, we cannot define an ordering amongst these nodes. Furthermore, if the strongly connected component is found near the top of the overall hierarchy, its effect is more severe compared to a situation where it is occurring only at the deeper levels. 

Finally, the orderability, $O$ is simply the fraction of nodes not taking part in any directed cycles, 
\begin{equation}
O=N_{\rm ac}/N,
\end{equation} where $N_{\rm ac}$ is the number of nodes that are not members in any directed cycles, and $N$ is the total number of nodes. Thus, orderability is analogous to the link-flow hierarchy $H_{\rm LF}$ given in (\ref{eq:linkflow}), the only difference is that here we measure the weight of the cycles in the network by the number of contained nodes instead of the number of contained links. Similarly to $H_{\rm LF}$, the oredrability is $O=1$ for all acyclic networks, while $O=0$ when the network is corresponding to a single directed cycle.

\section{Intensiveness}
We have seen on Fig.1.\ in the main paper that $H$ of finite regular trees is converging to the limit value for infinitely large trees already around $N=1000$ when $\lambda=2$, while the convergence is somewhat slower when $\lambda=4$. Nevertheless, above a certain $N$ it is the structure of the tree, (encoded in the branching number), what determines the hierarchy measure, not the size of the tree. This indicates that $H$ is behaving similarly to intensive quantities in physics in some aspects. 

To investigate this property further, let us examine what happens to $H$ if we take a pair of disjoint graphs ${\mathcal G}_1$ and ${\mathcal G}_2$, and unite them into a single graph ${\mathcal G}_1\cup {\mathcal G}_2$ with two isolated components. According to Eq.(10) in the main paper, the hierarchy measure of the graphs when considered separately can be written as
\begin{eqnarray}
H_{{\mathcal G}_1}&=&\sqrt{N_1\sum_{i=1}^{N_1}(p_i^{\rm stat})^2-1}, \label{eq:H_G_1}\\
H_{{\mathcal G}_2}&=&\sqrt{N_2\sum_{j=1}^{N_2}(p_j^{\rm stat})^2-1},  \label{eq:H_G_2}
\end{eqnarray}
where $p_i^{\rm stat}$ and $p_j^{\rm stat}$ correspond to the stationary distribution of the random walkers on nodes $i\in{\mathcal G}_1$ and $j\in{\mathcal G}_2$, while $N_1$ and $N_2$ denote the sizes of ${\mathcal G}_1$ and ${\mathcal G}_2$ respectively. When considering the union of the two graphs, the stationary distribution of the random walkers has to be normalised over the whole range of $N_1+N_2$ nodes. Thus, the stationary distribution on node $i$, originally in ${\mathcal G}_1$, now part of ${\mathcal G}_1\cup{\mathcal G}_2$, can be given as $\tilde{p}_i^{\rm stat}=p_i^{\rm stat}N_1/(N_1+N_2)$. Likewise, the  stationary distribution on node $j$, originally in ${\mathcal G}_2$, now part of ${\mathcal G}_1\cup{\mathcal G}_2$, can be expressed as $\tilde{p}_j^{\rm stat}=p_j^{\rm stat}N_2/(N_1+N_2)$. Therefore, the hierarchy measure of the union of the two graphs reads
\begin{eqnarray}
H_{{\mathcal G}_1\cup{\mathcal G}_2}&=&\sqrt{(N_1+N_2)\left(\sum_{i=1}^{N_1}(\tilde{p}_i^{\rm stat})^2 +\sum_{j=1}^{N_2}(\tilde{p}_j^{\rm stat})^2\right)-1}=\nonumber \\
& &\sqrt{\frac{N_1^2\sum\limits_{i=1}^{N_1}(p_i^{\rm stat})^2 +N_2^2\sum\limits_{j=1}^{N_2}(p_j^{\rm stat})^2}{N_1+N_2}-1}.
\label{eq:H_G_U_mid}
\end{eqnarray}
By rearranging (\ref{eq:H_G_1}-\ref{eq:H_G_2}) we gain 
\begin{eqnarray}
\sum_{i=1}^{N_1}(p_i^{\rm stat})^2&=&\frac{\left[H_{{\mathcal G}_1}\right]^2+1}{N_1}, \label{eq:p_1_sqr}\\ 
\sum_{i=1}^{N_1}(p_i^{\rm stat})^2&=&\frac{\left[H_{{\mathcal G}_1}\right]^2+1}{N_2}, \label{eq:p_2_sqr}
\end{eqnarray}
and by substituting (\ref{eq:p_1_sqr}-\ref{eq:p_2_sqr}) into (\ref{eq:H_G_U_mid}) we obtain
\begin{equation}
H_{{\mathcal G}_1\cup{\mathcal G}_2}=\sqrt{\frac{N_1\left[H_{{\mathcal G}_1}\right]^2+N_2\left[H_{{\mathcal G}_1}\right]^2}{N_1+N_2}}.
\label{eq:H_G_uni}
\end{equation}

Thus, the hierarchy measure of the union of two isolated networks is simply the weighted quadratic mean of $H$ obtained for the individual graphs. A noteworthy consequence is that if $H_{{\mathcal G}_1}=H_{{\mathcal G}_2}$, then this implies also that the hierarchy measure of the union is also the same, $H_{{\mathcal G}_1\cup{\mathcal G}_2}=H_{{\mathcal G}_1}=H_{{\mathcal G}_2}$. Furthermore, according to (\ref{eq:H_G_uni}) in general, if we take the union of ${\mathcal G}_1$ and ${\mathcal G}_2$ which are isolated from each other and $H_{{\mathcal G}_1}\leq H_{{\mathcal G}_2}$, then the hierarchy measure of the resulting network will be between the hierarchies measured for the two graphs separately, 
\begin{equation}
H_{{\mathcal G}_1}\leq  H_{{\mathcal G}_1\cup{\mathcal G}_2} \leq H_{{\mathcal G}_2}.
\end{equation}
Therefore, the rule for calculating $H$ for a system composed of isolated smaller parts based on $H$ obtained for these sub-systems is showing again a great deal of similarity to the behaviour of intensive physical quantities.

\section{The trees of maximal hierarchy}
\begin{figure}[hbt]
\centerline{\includegraphics[width=0.6\textwidth]{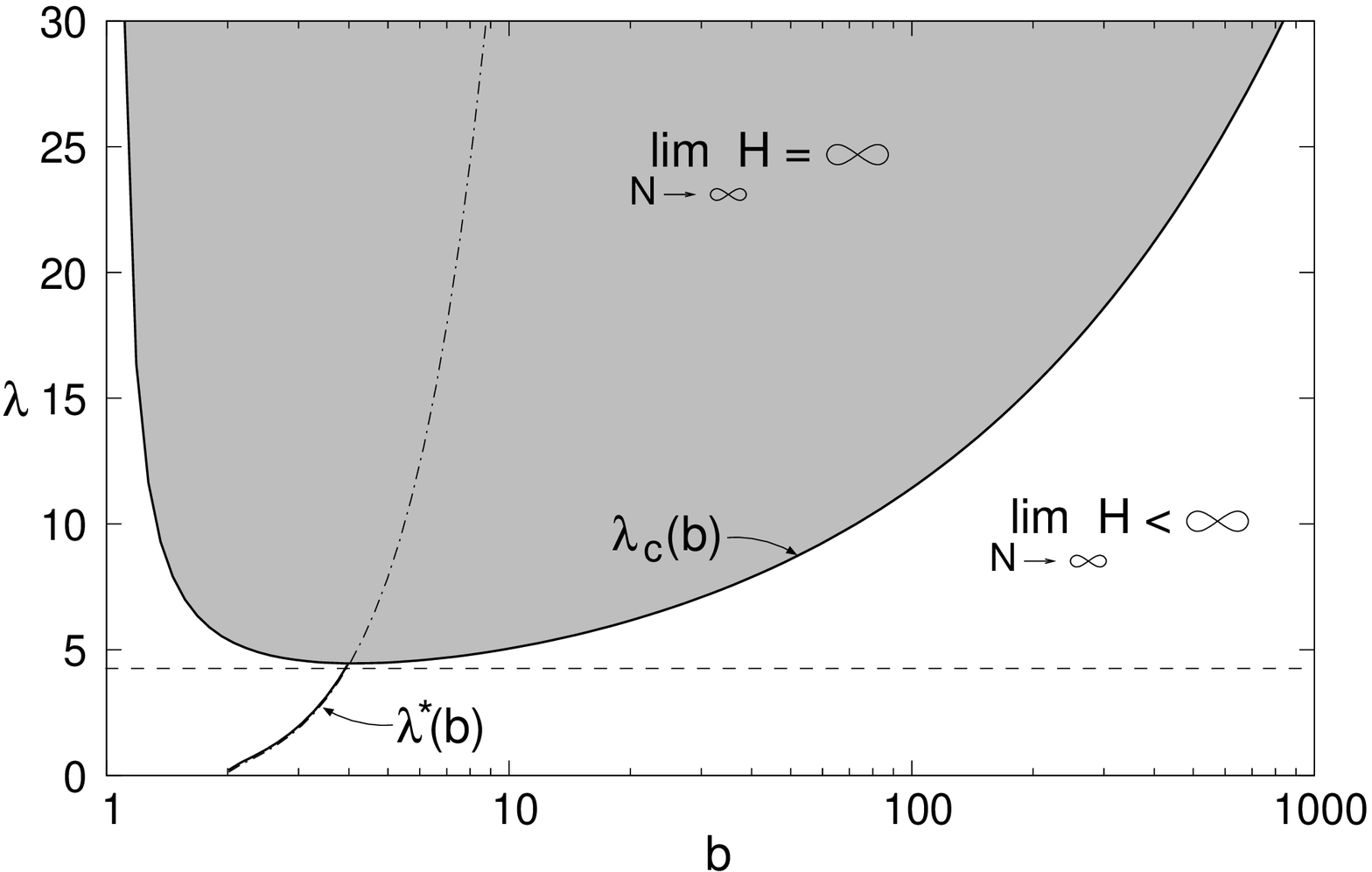}}
\caption{The ``phase diagram'' for $H$ in case of regular trees in the thermodynamic limit. If $\lambda$ and $b$ are falling into the region shown in grey, (meaning that $\lambda$ is larger than $\lambda_c(b)$ calculated from Eq.(23) in the main paper), then $H$ is diverging. In contrast, for parameter settings falling in the white region, $H$ remains finite in the thermodynamic limit. The maximal $H$ in this region is obtained at a $\lambda^*(b)$, given in (\ref{eq:lam_star}). The dashed horizontal line is showing the minimum of $\lambda_c(b)$.}
\label{fig:H_phase_diag}
\end{figure}

According to our results in the main paper, when considering regular trees of infinitely large size, the hierarchy measure may either diverge or remain finite, depending on both the branching number $b$ and the value of the $\lambda$ parameter. At a fixed $b$, the critical $\lambda_c(b)$ separating the two regimes is given by Eq.(23) in the main paper. Here, in Fig.\ref{fig:H_phase_diag}.\ we show $\lambda_c$ as a function of $b$ on a semi-logarithmic plot, where the regime of diverging $H$ values is coloured grey. 

\begin{figure}[hbt]
\centerline{\includegraphics[width=0.6\textwidth]{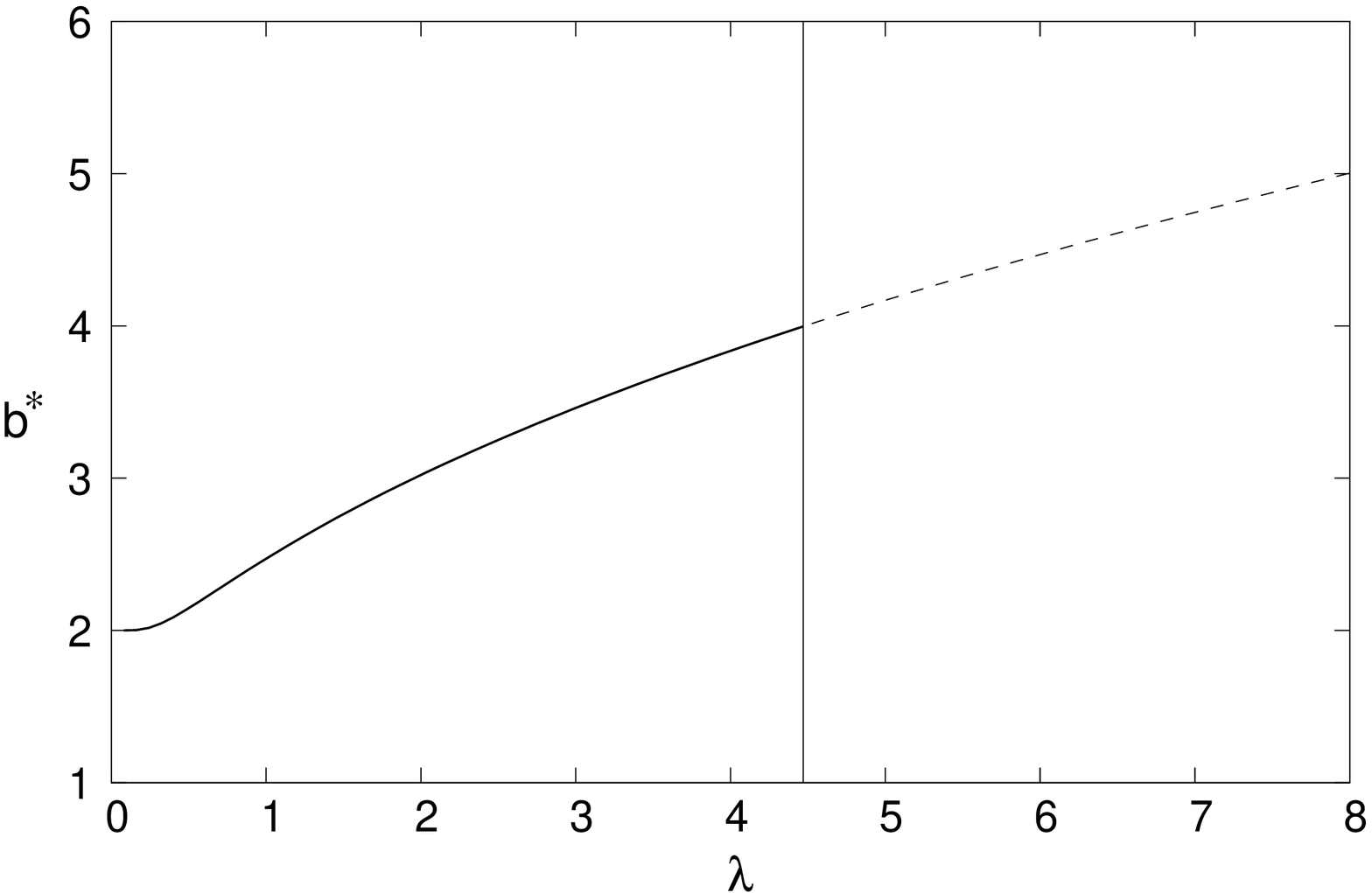}}
\caption{The branching number $b^*$ for which $H$ is maximal, as a function of the parameter $\lambda$ in case of infinitely large regular trees. The vertical line is corresponding to the minimum of $\lambda_c(b)$ given in Eq.(23) in the main paper, thus, for $\lambda$ values above that $H$ can become divergent in the thermodynamic limit.}
\label{fig:max_helyek}
\end{figure}

In the regime of finite $H$, the limit value for the hierarchy measure is given by Eq.(21) in the main paper. The maximum of this function can be located by solving 
\begin{equation}
b^2(e^{1/\lambda}-1)-2b(e^{1/\lambda}-1)-2=0.
\end{equation}
Based on that, at a fixed branching number $b$, the $\lambda$ parameter providing the maximal $H$ can be written as
\begin{equation}
\lambda^*=\left[\ln\left(\frac{2}{b^2-2b}+1\right)\right]^{-1},
\label{eq:lam_star}
\end{equation}
while at a fixed $\lambda$ parameter the tree with maximal hierarchy has a branching number of
\begin{equation}
b^*=1+\frac{\sqrt{e^{2/\lambda}-1}}{e^{1/\lambda}-1}.
\label{eq:b_star}
\end{equation}
In Fig.\ref{fig:H_phase_diag}. we also show $\lambda^*$ given in (\ref{eq:lam_star}) as a function of $b$. In parallel, the $b^*$ expressed in (\ref{eq:b_star}) is plotted as a function of $\lambda$ in Fig.\ref{fig:max_helyek}. 
According to picture, the maximum of $H$ is at $b=2$ when $\lambda$ is low. However, $b^*$ is steadily increasing as a function of $\lambda$, reaching to $b^*\approx 3.84$ at $\lambda=4$, chosen to be the optimal $\lambda$ value based on arguments given in the main paper.

\section{Hierarchy of real networks}
We have analysed the hierarchy in a large number of networks of different types and varying sizes, the scatter plot of the obtained $H$ values and the $z$-score of $H$ are given in Figs.5-6.\ in the main paper. In Tables \ref{tab:results_real_prop}-\ref{tab:results_real_rand}.\ we provide the results in more details by listing also the size and the average degree beside the $H$ values and $z$-scores of the examined networks. 

\begin{table}[ht!] 
\centering
\tabcolsep=0.09cm
\footnotesize
\rowcolors{1}{lightgray}{white}
\begin{tabular}{|c|c|c|c|c|c|}
\hline
Type & Meaning of $A\to B$ & Network & $N$ & $\langle k \rangle=\frac{M}{N}$ & $H$ \\
\hline

Electric & $B$ depends on& s1488 \cite{ElectricNetworkData} & $667$ & $2.085$ & $0.893$ \\
&the value at $A$ & s1494 \cite{ElectricNetworkData} & $661$ & $2.116$ & $0.880$\\
&& s5378 \cite{ElectricNetworkData} & $2993$ & $1.467$ & $0.887$ \\
&& s9234 \cite{ElectricNetworkData} & $5844$ & $1.400$ & $0.870$ \\
&& s35932 \cite{ElectricNetworkData} & $17828$ & $1.683$ & $0.719$ \\

Citation & $A$ is cited by $B$ & ArXiv-HepPh \cite{leskovec2005graphs} & $34546$ & $12.203$ & $0.352$\\
&& ArXiv-HepTh \cite{leskovec2005graphs} & $27770$ & $12.705$ & $0.430$\\

Food web & $A$ eats $B$ & GrassLand \cite{dunne2002food} & $88$ & $1.557$ & $1.551$\\
&& LittleRock \cite{martinez1991artifacts} & $183$ & $13.628$ & $0.482$\\
&& St.\ Marks \cite{christian1999organizing} & $49$ & $4.612$ & $0.634$\\
&& Ythan \cite{dunne2002food} & $135$ & $4.452$ & $0.677$\\

Internet & $A$ sent messages & p2p-1 \cite{leskovec2007graph} & $10876$ & $3.677$ & $0.524$\\
& to $B$& p2p-2 \cite{leskovec2007graph} & $8846$ & $3.599$ & $0.541$\\
&& p2p-3 \cite{leskovec2007graph} & $8717$ & $3.616$ & $0.529$\\

Metabolic & $B$ is an end & C. elegans \cite{jeong2000large} & $1173$ & $2.442$ & $0.467$\\
&product of $A$ & E. coli \cite{jeong2000large} & $2275$ & $2.533$ & $0.447$\\
&& S. cerevisiae \cite{jeong2000large} & $1511$ & $2.537$ & $0.440$\\

Organization & $B$ trusts in $A$ & Consulting \cite{cross2004hidden} & $46$ & $19.109$ & $0.120$\\
&& Enron \cite{klimt2004introducing} & $156$ & $10.699$ & $0.126$\\
&& Manufacturing \cite{cross2004hidden} & $34$ & $18.935$ & $2.18\times 10^{-2}$\\
& $B$ knows $A$ & Freemans-1 \cite{freemans} & $34$ & $18.971$ & $6.35\times 10^{-2}$\\
&& Freemans-2 \cite{freemans} & $77$ & $24.412$ & $1.06\times 10^{-2}$\\

Regulatory & $A$ regulates $B$ & TRN-Yeast-1 \cite{balaji2006comprehensive} & $4441$ & $2.899$ & $0.262$\\
&& TRN-Yeast-2 \cite{milo2002network} & $688$ & $1.568$ & $0.720$\\
&& TRN-EC \cite{milo2002network} & $419$ & $1.239$ & $0.823$\\

Trust & $B$ trusts in $A$ & College \cite{van2003evolution} & $32$ & $3.000$ & $0.464$\\
&& Epinions \cite{richardson2003trust} & $75888$ & $6.705$ & $0.437$\\
&& Prison \cite{milo2004superfamilies} & $67$ & $2.716$ & $0.565$\\
&& WikiVote \cite{leskovec2010signed}& $7115$ & $14.573$ & $0.201$\\

Language & word $B$ & English \cite{i2001small} & $7724$ & $5.992$ & $0.404$\\
&follows word $A$ & French \cite{i2001small}& $9424$ & $2.578$ & $0.478$\\
&& Spanish \cite{i2001small}& $12642$ & $3.570$ & $0.194$\\
&& Japanese \cite{i2001small}& $3177$ & $2.613$ & $0.497$\\

World Wide& $B$ has a link to $A$ & Google web \cite{palla2007directed} & $15763$ & $10.861$ & $0.258$\\
 Web & & nd.edu \cite{albert1999internet} & $325729$ & $4.596$ & $0.557$\\
&& Polblogs \cite{adamic2005political} & $1490$ & $12.812$ & $0.223$\\

\hline
\end{tabular}
\caption{Random walk hierarchy of real networks shown in Fig.5.\ in the main paper. The network type is given in the 1$^{\rm st}$ column, the meaning of the links in the $2^{\rm nd}$ column, the references to the data sources are listed in the $3^{\rm rd}$ column. The network size is given in the 4$^{\rm th}$ column, followed by the average degree in the 5$^{\rm th}$ column. The hierarchy measure $H$ (calculated at the optimal $\lambda=4$ parameter value) is provided in the 6$^{\rm th}$ column.}
\label{tab:results_real_prop}
\end{table}

The electric networks, where the target of a directed link is depending on the value of the source node, turned out to be rather hierarchical according to our measure. Although their sizes is varying between $N={\mathcal O}(10^2)$ and $N={\mathcal O}(10^4)$, the hierarchy values are quite close to each other, forming an  elongated cluster in Fig.5.\ in the main paper. The analysed two citation networks, where the links are pointing from the cited paper to the citing article, showed only a moderate amount of hierarchy. In contrast, a part of the food webs were amongst the most hierarchical networks in the studied examples. However, the $H$ values showed a relative large variance for this network type, which can be due to the different habitat of the species involved in the listed food webs. Furthermore, in case of the St. Marks food web the links were weighted, with the weights corresponding to the fractions of the consumer’s diet, whereas in the other cases the links were un-weighted. 

\begin{table}[ht!] 
\centering
\tabcolsep=0.09cm
\footnotesize
\rowcolors{1}{lightgray}{white}
\begin{tabular}{|c|c|c|c|c|}
\hline
Type & Network & $H$ & $\left< H_{\rm rand}\right> \pm \sigma(H_{\rm rand})$ & $z$\\
\hline

Electric & s1488 \cite{ElectricNetworkData} & $0.893$ & $0.811\pm 1.66\cdot 10^{-2}$ & $4.94$ \\
         & s1494 \cite{ElectricNetworkData} & $0.880$ & $0.780\pm 1.83\cdot 10^{-2}$ & $5.46$ \\
         & s5378 \cite{ElectricNetworkData} & $0.887$ & $0.803\pm 1.65\cdot 10^{-2}$ & $5.09$ \\
         & s9234 \cite{ElectricNetworkData} & $0.870$ & $0.861\pm 1.23\cdot 10^{-2}$ & $0.73$ \\
         & s35932 \cite{ElectricNetworkData}& $0.719$ & $0.787\pm 5.44\cdot 10^{-3}$ & $-12.5$ \\

Citation & ArXiv-HepPh \cite{leskovec2005graphs} & $0.352$ & $0.222\pm 5.16\cdot 10^{-4}$ & $251.94$ \\
         & ArXiv-HepTh \cite{leskovec2005graphs} & $0.430$ & $0.221\pm 2.46\cdot 10^{-3}$ & $84.96$ \\

Food web & GrassLand \cite{dunne2002food} & $1.551$ & $1.036\pm 6.20\cdot 10^{-2}$ & $8.31$ \\
         & LittleRock \cite{martinez1991artifacts} & $0.482$ & $0.221\pm 1.68\cdot 10^{-2}$ & $15.54$  \\
         & St.\ Marks \cite{christian1999organizing} & $0.634$ &  $0.425\pm 3.44\cdot 10^{-2}$ & $6.08$ \\
         & Ythan \cite{dunne2002food} & $0.677$ & $0.450\pm 2.19\cdot 10^{-2}$ & $10.37$ \\

Internet & p2p-1 \cite{leskovec2007graph} & $0.524$ & $0.488\pm 1.48\cdot 10^{-3}$ & $24.32$ \\
         & p2p-2 \cite{leskovec2007graph} & $0.541$ & $0.499\pm 1.62\cdot 10^{-3}$ & $25.92$ \\
         & p2p-3 \cite{leskovec2007graph} & $0.529$ & $0.495\pm 1.61\cdot 10^{-3}$ & $21.12$  \\

Metabolic & C. elegans \cite{jeong2000large} & $0.467$ & $0.611\pm 1.20\cdot 10^{-2}$ & $-12.00$ \\
          & E. coli \cite{jeong2000large} & $0.447$ & $0.630\pm 8.99\cdot 10^{-3}$ & $-20.36$ \\
          & S. cerevisiae \cite{jeong2000large} & $0.440$ & $0.609\pm 1.14\cdot 10^{-2}$ & $-14.82$ \\

Organization & Consulting \cite{cross2004hidden} & $0.120$ & $4.99\cdot 10^{-2}\pm 5.24\cdot 10^{-3}$ & $13.38$ \\
             & Enron \cite{klimt2004introducing} & $0.126$ & $0.160\pm 1.44\cdot 10^{-2}$ & $-2.36$ \\
             & Manufacturing \cite{cross2004hidden} & $2.18\times 10^{-2}$ & $2.04\cdot 10^{-2} \pm 1.27\cdot 10^{-3}$ & $1.10$ \\
             & Freemans-1 \cite{freemans} & $6.35\times 10^{-2}$ & $6.01\cdot 10^{-2}\pm 1.72\cdot 10^{-3}$ & $1.98$ \\
             & Freemans-2 \cite{freemans} & $1.06\times 10^{-2}$ & $9.42\cdot 10^{-3}\pm 1.20\cdot 10^{-3}$ & $0.98$ \\

Regulatory & TRN-Yeast-1 \cite{balaji2006comprehensive} & $0.262$ & $0.264\pm 2.39\cdot 10^{-3}$ & $-0.84$ \\
           & TRN-Yeast-2 \cite{milo2002network} & $0.720$ & $0.733\pm6.45\cdot 10^{-3}$ & $-2.02$ \\
           & TRN-EC \cite{milo2002network} & $0.823$ & $0.853\pm8.22\cdot 10^{-3}$ & $-3.65$ \\

Trust & College \cite{van2003evolution} & $0.464$ & $0.403\pm 2.82\cdot 10^{-2}$ & $2.16$ \\
      & Epinions \cite{richardson2003trust} & $0.437$ & $0.373\pm 5.15\cdot 10^{-3}$ & $12.43$ \\
      & Prison \cite{milo2004superfamilies} & $0.565$ & $0.565\pm 4.94\cdot 10^{-2}$ & $0$ \\
      & WikiVote \cite{leskovec2010signed}& $0.201$ & $0.157\pm 1.49\cdot 10^{-3}$ & $29.53$ \\

Language & English \cite{i2001small} & $0.404$ & $0.388\pm 4.89\cdot 10^{-3}$ & $3.27$ \\
         & French \cite{i2001small}& $0.478$ & $0.519\pm 4.64\cdot 10^{-3}$ & $-8.83$ \\
         & Spanish \cite{i2001small}& $0.194$ & $0.241\pm 4.03\cdot 10^{-3}$ & $-11.66$ \\
         & Japanese \cite{i2001small}& $0.497$ & $0.560\pm 8.24\cdot 10^{-3}$ & $-7.65$ \\

World Wide& Google web \cite{palla2007directed} & $0.258$ & $0.246\pm 1.68\cdot 10^{-3}$ & $7.14$ \\
          & nd.edu \cite{albert1999internet} & $0.557$ & $0.451 \pm 7.84\cdot 10^{-4}$ & $135.20$ \\
          & Polblogs \cite{adamic2005political} & $0.223$ & $0.184\pm 8.21\cdot 10^{-3}$ & $4.75$ \\

\hline
\end{tabular}
\caption{Comparing $H$ in real networks to that of their link randomised counterparts. The network type is given in the $1^{\rm st}$ column, the references to the data sources are listed in the $2^{\rm nd}$ column, and the $H$ measured in the original networks is given in the $3^{\rm rd}$ column. The average value of $H$ in the configuration model, (calculated via link randomisation) is presented in the $4^{\rm th}$ column, accompanied by the standard deviation. Finally, the corresponding $z$-scores are listed in the $5^{\rm th}$ column.}
\label{tab:results_real_rand}
\end{table}

The peer to peer message networks over the Internet and the metabolic networks, (where the target of a link was corresponding to a product of the source) showed  moderate levels of hierarchy. For both types, the data points in Fig.5.\ in the main paper formed tight clusters. The intra organisational networks were the least hierarchical according to our measure, with the data points forming a  cluster close to the origin. Similarly to the food webs, the variance of the $H$ values obtained for the regulatory networks was relatively high. I.e., the regulatory networks TRN-Yeast-2 and TRN-Yeast-1  are amongst the most hierarchical studied systems, whereas the TRN-EC network shows only a modest hierarchy value. 

The trust networks obtained moderate hierarchy values, and the variance of their $H$ values is far smaller compared to the very large variance in their sizes. The language networks turned out to be moderately hierarchical as well. Here the networks were originating from large text corpora, and a directed link from word $A$ to $B$ is signaled that $B$ was following $A$ in the  text. According to the results, the English- French- and Japanese language networks obtained $H$ values quite close to each other, while the hierarchy of the Spanish language seemed considerably lower. Finally, the networks between web pages showed again a moderate hierarchy, where $H$ for the Google web and Polblogs were quite close to each other, while network of nd.edu received a significantly higher hierarchy value.

It is very interesting to see how the picture is changing when we switch from $H$ to the $z$-score shown in Fig.6 in the main paper, corresponding to the difference between $H$ and $\left< H\right>$ in the configuration model, scaled by standard deviation of $H$ in the configuration model. The citation networks, the Internet networks and the food webs are forming three rather tight clusters in Fig.6. in the main paper with significant positive $z$-score, (which is outstandingly high in case of the citation networks). In contrast, the metabolic- and regulatory networks are forming clusters of significant negative $z$-score, indicating that their structure is far less hierarchical compared to what we would expect by assuming random connections between the nodes, (at the same degree distribution as in the original network). The $z$-scores of WWW networks were all positive, (where the nd.edu network showed an outstandingly high value). In case of the trust networks the $z$-scores are also almost always positive, with the exception of the Prison network, where $z=0$. The organisational-, electric-, and language networks showed a mixed picture, including both positive and negative $z$-scores.

\end{document}